\documentclass[default,iicol]{sn-jnl}% Default with double column layout

\usepackage{graphicx}%
\usepackage{multirow}%
\usepackage{amsmath,amssymb,amsfonts}%
\usepackage{amsthm}%
\usepackage{mathrsfs}%
\usepackage[title]{appendix}%
\usepackage{xcolor}%
\usepackage{textcomp}%
\usepackage{manyfoot}%
\usepackage{booktabs}%
\usepackage{algorithm}%
\usepackage{algorithmicx}%
\usepackage{algpseudocode}%
\usepackage{listings}%
\usepackage{dcolumn}% Align table columns on decimal point
\usepackage{bm}% bold math
\usepackage{float}
\usepackage{CJK}
\usepackage{array}
\usepackage{subfigure}
\usepackage{soul}
\usepackage{color}
\usepackage[justification=centering]{caption}
\begin{document}

\title[Article Title]{Virtual segmentation of a small contact HPGe detector:
inference of hit positions of single-site events via pulse shape analysis}

\author[1]{\fnm{W.H.} \sur{Dai}}
\author*[1]{\fnm{H.} \sur{Ma}}\email{mahao@tsinghua.edu.cn}
\author[1]{\fnm{Z.} \sur{Zeng}}
\author[1]{\fnm{L.T.} \sur{Yang}}
\author[1]{\fnm{Q.} \sur{Yue}}
\author[2]{\fnm{J.P.} \sur{Cheng}}

\affil[1]{
    \orgdiv{Key Laboratory of Particle and Radiation Imaging 
    (Ministry of Education) and Department of Engineering Physics}, 
    \orgname{Tsinghua University}, 
    \orgaddress{
    \city{Beijing}
    \postcode{100084}, 
    \country{China}}
}
\affil[2]{
    \orgname{Beijing Normal University}, 
    \orgaddress{
    \city{Beijing}
    \postcode{100875}, 
    \country{China}}
}


\abstract{
    Exploring hit positions of recorded events can help to
    understand and suppress backgrounds in rare event searching experiments. 
    In this study, we virtually segment a 
    small contact P-type high purity germanium detector (HPGe) into two layers. 
    Single-site events (SSEs) in each layer are selected by an algorithm
    based on two pulse shape parameters: the charge pulse drift time ($T_{Q}$)
    and current pulse rise time ($T_{I}$).
    To determine the shapes and volumes of the two layers,
    a Th-228 source is placed at top and side positions to irradiate the detector.
    The double escape peak events from 2614.5 keV $\gamma$-ray are selected 
    as typical SSEs, their numbers in the two layers are used to calculate
    the volumes and shapes of those layers.
    Considering the statistical and systematic uncertainties,
    the inner layer volume is evaluated to be 
    47.2\%$\pm$0.26(stat.)\%$\pm$0.22(sys.)\%
    of the total sensitive volume. 
    We extend our analysis for SSEs in 1400-2100 keV,
    the spectra of inner layer events acquired from experimental data using the selection algorithm
    are in good agreement with those from the simulation. 
    For sources outside the HPGe detector, 
    the outer layer can act as a shielding for the inner layer. 
    Selecting the inner layer as the analysis volume can reduce the external
    background in the signal region of Ge-76 neutrinoless double beta (0$\nu\beta\beta$) decay.
    We use the Th-228 source to evaluate the background suppression power of the virtual segmentation.
    After performing the single and multi-site event discrimination,
    the event rate in the 0$\nu\beta\beta$ signal region can be 
    further suppressed by 12\% by selecting the inner layer as the analysis volume.
    The virtual segmentation could be used to efficiently suppress surface background like
    electrons from Ar-42/K-42 decay in 0$\nu\beta\beta$ experiments using germanium detector immersed in liquid argon.
}

%%================================%%
%% Sample for structured abstract %%
%%================================%%


\maketitle

\section{Introduction}\label{sec1}

Small contact high purity germanium (HPGe) detectors are widely 
used in searching for rare events from physics beyond Standard Model, 
such as the neutrinoless double beta ($0\nu\beta\beta$) decay
and dark matter \cite{bib:4,bib:5,bib:6,bib:7}. 
Those searches need an extremely low background level in 
the signal region to achieve sufficient sensitivity.
The discrimination of background and signal via pulse shape analysis 
is a powerful background suppression technology and widely used in
HPGe based experiments.
\cite{bib:8,bib:9,bib:10,bib:11}.

The energy depositions from $0\nu\beta\beta$ decay events 
and dark matter interactions are typically within about a 
millimeter and are regarded as single-site events (SSEs). 
Backgrounds can be single-site or multi-site events (MSEs), 
depending on their origination. Small contact HPGe detectors, 
such as point contact Ge (PCGe) and broad energy Ge (BEGe), 
have been demonstrated to have SSE and MSE discrimination 
capability utilizing pulse shape analysis 
\cite{bib:3,bib:9,bib:10,bib:11}.
After the SSE/MSE discrimination, signals are still mixed 
with SSE-like backgrounds, such as single Compton scattering of 
incoming $\gamma$ or direct energy depositions from 
beta decay electrons penetrating the surface layer of the detector. 
Signals are expected to have a uniform distribution in the detector, 
while the backgrounds tend to be close to the detector surface. 
Therefore, inference of the SSE position can help to understand 
and suppress the SSE-like backgrounds.

Previous studies \cite{bib:12,bib:13,bib:14}
have demonstrated that the charge collection time in a small 
contact HPGe detector depends on the energy deposition position. 
Past work \cite{bib:13} has shown 
that the rise time of the event pulse can be used to estimate 
the distance of energy deposition from the contact in a PCGe detector. 
Pulse shape simulation in \cite{bib:12}
also showed that the signal shape depends on the interaction position.

This work explores the position discrimination power of a small contact 
$p$-type HPGe detector via pulse shape analysis. 
The detector is virtually segmented into two layers, and single-site 
events with hit position in the inner layer are identified. 
The shape and volume of the inner layer are 
modeled, determined, and validated in a series of Th-228 irradiation experiments. 
We also discuss the background suppression potential of this method 
towards possible application in future $0\nu\beta\beta$ experiments.

\begin{figure}[!htb]
    \centering
    \includegraphics
    [width=1.0\hsize]
    {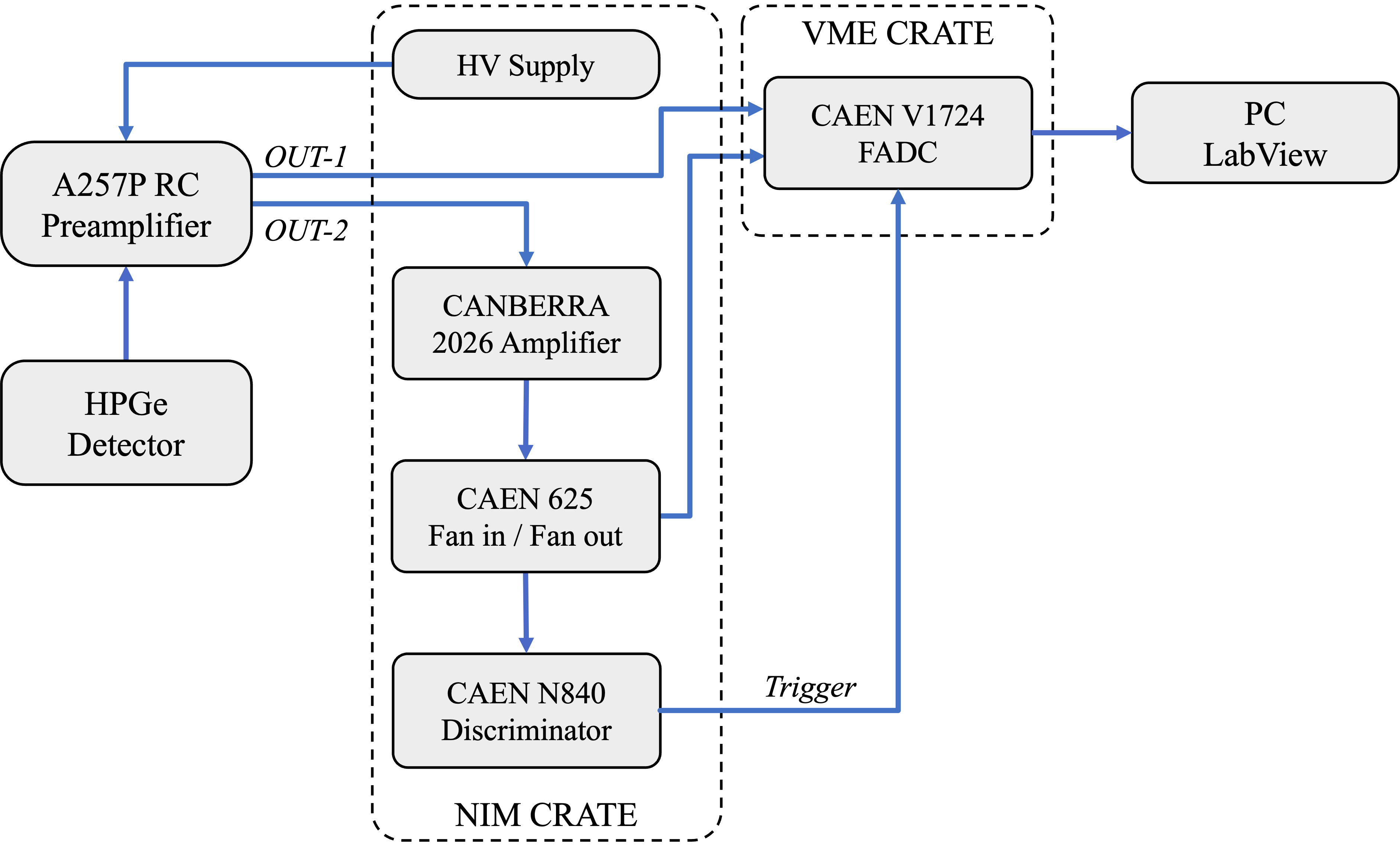}
    \caption{\label{fig:DAQ} Schematic diagram of the DAQ system.}
\end{figure}

\begin{figure}[!htb]
    \centering
    \includegraphics
    [width=0.65\hsize]
    {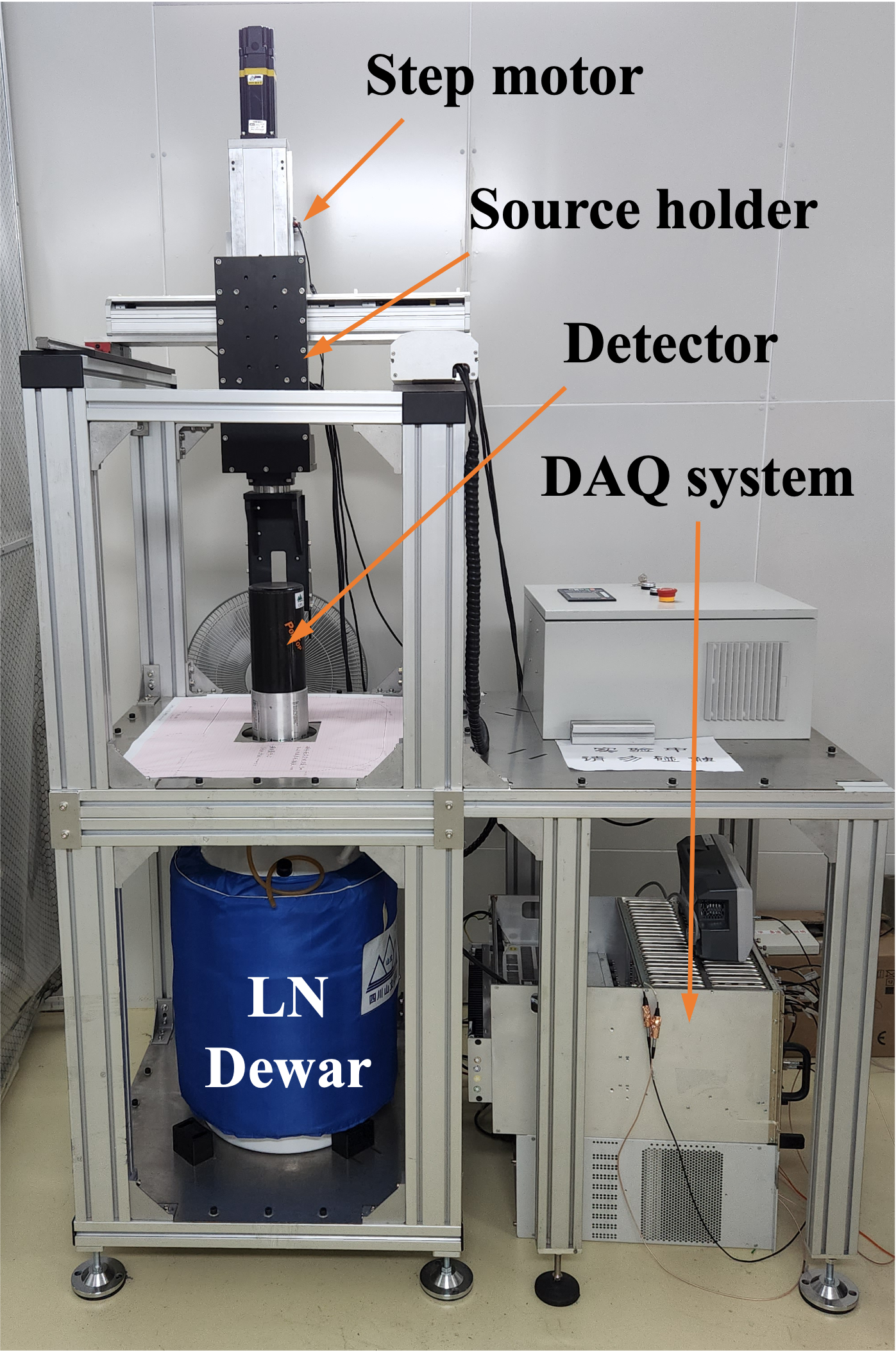}
    \caption{\label{fig:setup} Experimental setup at CJPL.}
\end{figure}

\begin{figure*}[htbp]
    \centering
    \includegraphics
    [width=1.0\hsize]
    {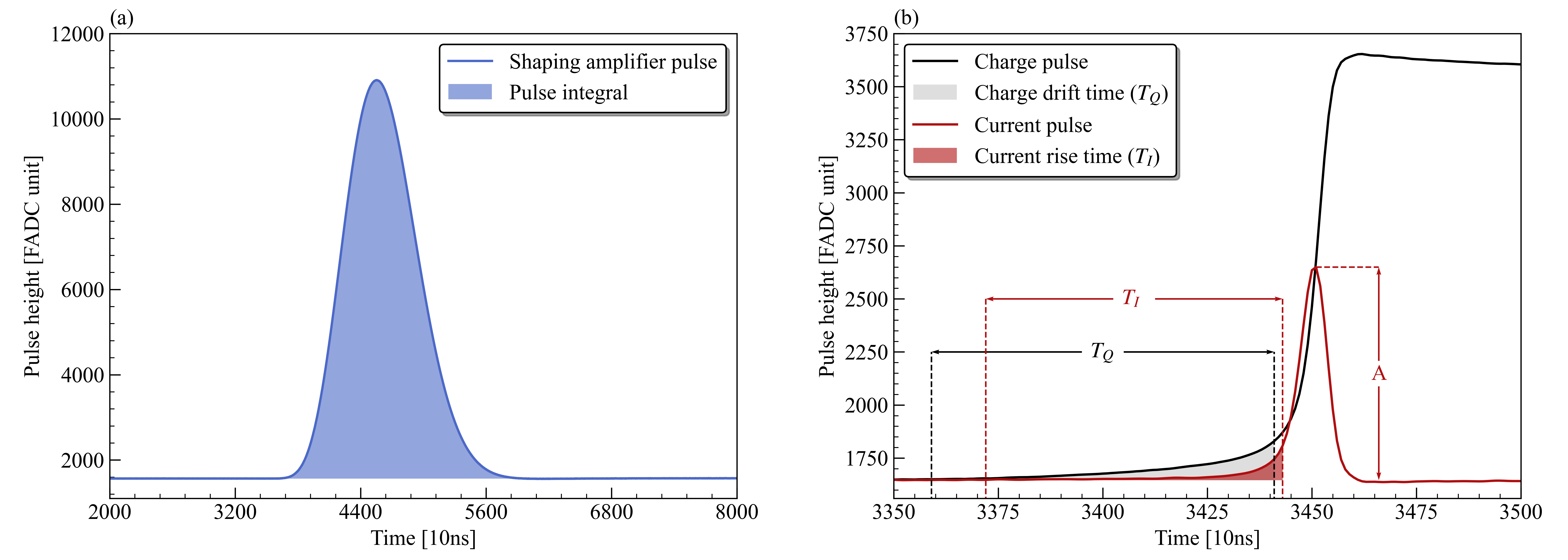}
    \caption{\label{fig:dpp} 
    (a) an example of shaping amplifier pulse, the blue region indicates 
    the integral of the pulse after subtracting the baseline, and it is 
    used as the energy estimator; 
    (b) an example of smoothed preamplifier pulse and the extracted 
    current pulse. Pulse time parameters $T_Q$, $T_I$, and parameter 
    "A" in the A/E discriminator are also illustrated. 
    The current pulse is rescaled for demonstration.}
\end{figure*}

\section{Experimental setup}\label{sec2}
The detector used in this work is a small contact $p$-type HPGe 
detector produced by ORTEC. 
The detector crystal has a height of 42.6 mm and a diameter of 80.0 mm, 
and the thin $p$+ contact is about 3.1 mm in diameter and is
implemented in a 1 mm deep hole on the bottom surface of the crystal. 
The $n$+ surface of the detector crystal, formed by the lithium diffusion, 
contains an inactive layer and reduces the sensitive mass of the detector.
The thickness of the inactive layer is evaluated to be 0.87 mm 
in our previous work \cite{bib:15}. Subtracting the inactive layer,
the total sensitive mass of the detector is 1.052 kg.

As shown in Fig.\ref{fig:DAQ}, the data acquisition (DAQ) system 
is based on commercial NIM/VME modules and crates. 
The detector is operated under 4500 V bias voltage provided by a high 
voltage module. The output signal from the $p$+ contact 
is fed into an resistance-capacitance (RC) preamplifier. 
The RC-preamplifier provides two identical output signals. One is loaded 
into a shaping amplifier with a gain factor of 10 and 
shaping time of 6 $\mu$s. The output of the shaping amplifier and the other output of 
the RC-preamplifier are fed into a 14-bit 100 MHz flash analog-to-digital 
convertor (FADC) for digitalization. The digitalized waveforms 
are recorded by the DAQ software on a PC platform. 

A detector scanning device is built in China Jinping 
Underground Laboratory (CJPL) \cite{bib:16}. 
As shown in Fig.\ref{fig:setup}, 
the detector and the liquid nitrogen (LN) Dewar are installed with the scanning device.
A Th-228 source with an activity of 500 Bq is mounted 
on the source holder with a step motor controlling the source position.

\section{Pulse processing and event discrimination}\label{sec3}
\subsection{Digital pulse processing}\label{sec3.1}

Typical pulses from the shaping amplifier and preamplifier are illustrated 
in Fig.\ref{fig:dpp}. After subtracting the baseline, the integration of the shaping 
amplifier pulse is used to estimate the event energy (as shown in Fig.\ref{fig:dpp}(a)). 
Energy calibration is performed by the measured Th-228 spectrum with characteristic 
$gamma$-ray peaks from decays of radionuclides in the Th-228 decay chain.

The pulses from the preamplifier are used to estimate the time features of the event 
(as shown in Fig.\ref{fig:dpp}(b)). The charge drift time ($T_Q$) is defined as the 
time between the moments when charge pulse reachs 0.2\% and 10\% of its maximum amplitude. 
The current pulse is extracted from the charge pulse by a moving average 
differential filter, and the current rise time ($T_I$) is the time between the moments when 
the current pulse reachs 0.2\% and 20\% of its maximum amplitude.

\subsection{Single and multi-site event discrimination}\label{sec3.2}

\begin{figure}[!htb]
    \centering
    \includegraphics
    [width=1.0\hsize]
    {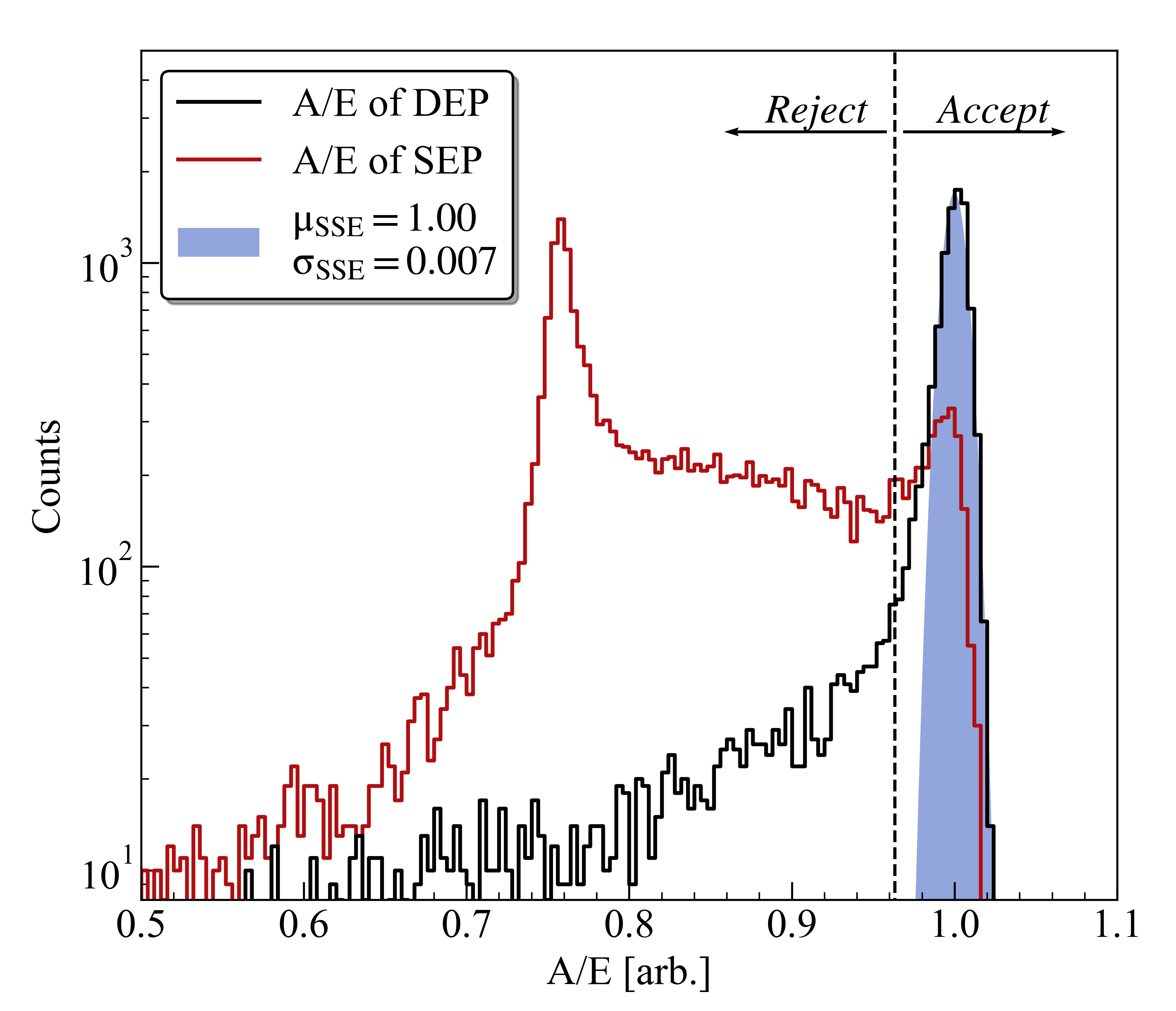}
    \caption{\label{fig:AE1} A/E distributions of DEP and SEP events in 
    Th-228 calibration data. The dashed line is the A/E cut threshold 
    ($\mu_{SSE}-5\sigma_{SSE}$).}
\end{figure}

The single/multi-site event discriminator (A/E) is defined as
ratio of the maximum amplitude of the current 
pulse (A) and the reconstructed energy (E). 
It has been discussed in various literature 
\cite{bib:9,bib:11,bib:17,bib:18}
that SSE tends to have higher A/E value than MSE in a small contact HPGe detector. 
Therefore, we apply a cut on A/E to select the SSEs. The acceptance region of the A/E cut 
is determined by the double escape peak (DEP) events from a measured Th-228 spectrum. 
DEP events are typical SSEs and their A/E distribution is fitted by a Gaussian function 
to determine the mean ($\mu_{SSE}$) and standard deviation ($\sigma_{SSE}$) of A/E parameter for SSEs. 
As shown in Fig.\ref{fig:AE1}, 
the cut threshold is set to $\mu_{SSE}-5\sigma_{SSE}$, leading to about 80\% survival 
fraction of DEP events and 9\% survival fraction of single escape peak events (typical MSEs). 

\begin{figure}[!htb]
    \centering
    \includegraphics
    [width=1.0\hsize]
    {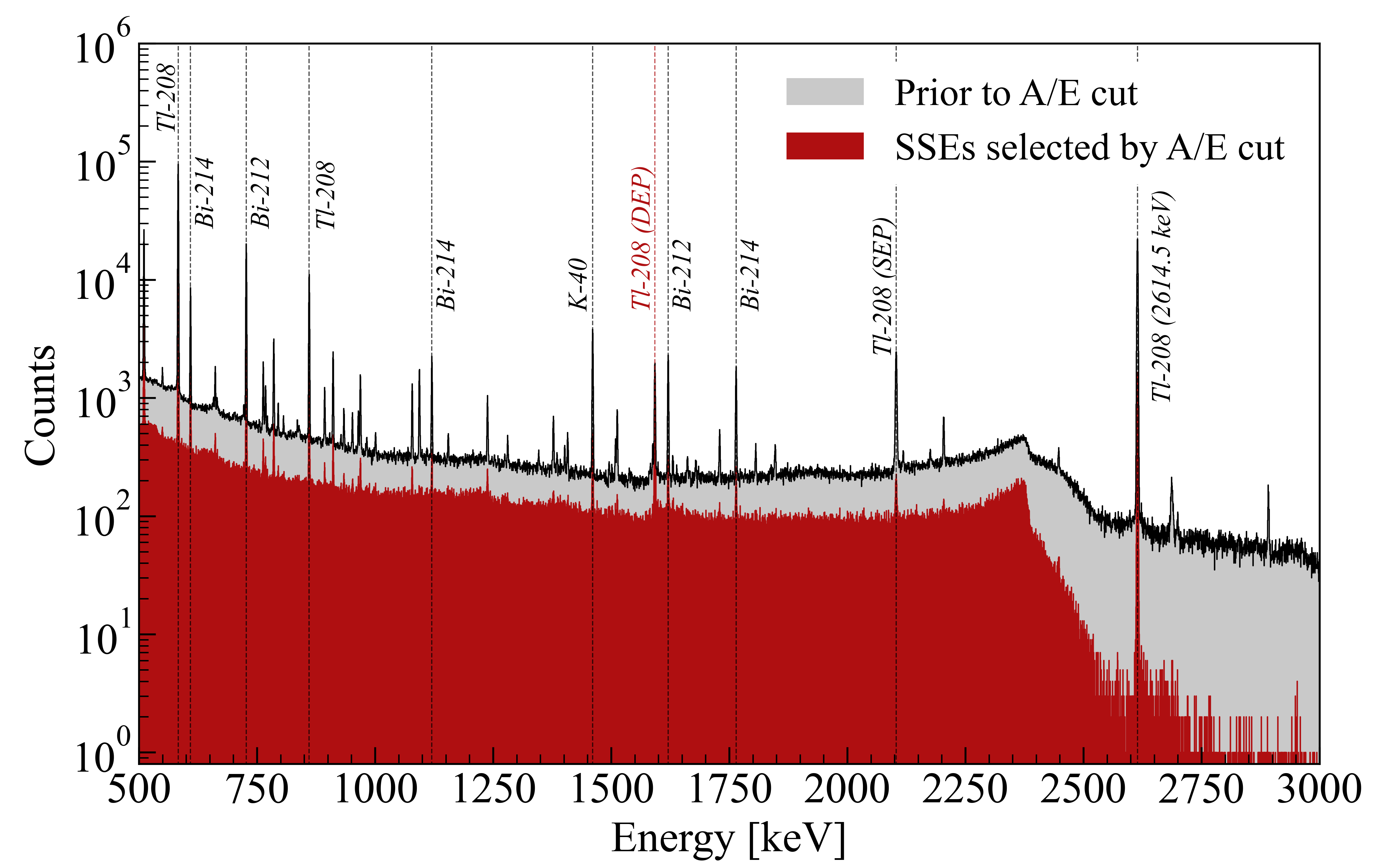}
    \caption{\label{fig:AE2} 
    Typical Th-228 spectra before and after the A/E cut.
    The characteristic peaks from decay daughters of Th-228
    (Tl-208, Bi-212) and other radionuclides (K-40, and Bi-212)
    are labeled in the spectra. The double-escape peak (DEP) of Tl-208
    2614.5 keV $\gamma$-ray is marked in red.
    }
\end{figure}

Fig.\ref{fig:AE2} shows typical Th-228 spectra before and after the A/E cut.
Main characteristic peaks from the Th-228 source and radionuclides in the 
surrounding materials are labeled.
The full-width-at-half-maximum (FWHM) of the double escape peak (1592.5 keV)
before (after) the A/E cut is $2.19\pm 0.05$ keV ($2.18\pm 0.03$ keV).
The FWHM of the 2614.5 keV peak before (after) the A/E cut is
$2.51\pm 0.01$ keV ($2.46\pm 0.02$ keV).
A slight improvement in the energy resolution is observed after the A/E cut.

\subsection{Linear and nonlinear event discrimination}\label{sec3.3}

\begin{figure*}[htbp]
    \centering
    \includegraphics
    [width=1.0\hsize]
    {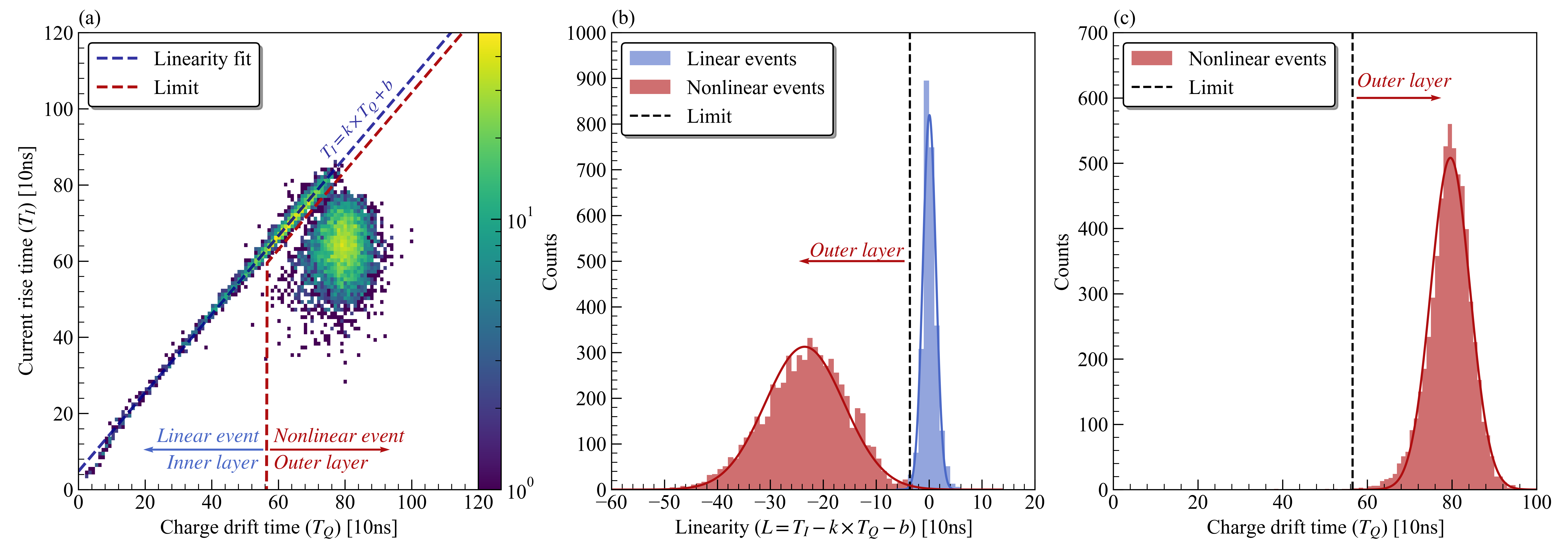}
    \caption{\label{fig:linearity} 
    Discrimination of linear and nonlinear events. 
    Data in the figure are from DEP events 
    (1592.5$\pm$5 keV, after A/E cut) in a Th-228 calibration 
    experiment (source placed at the center of detector top surface). 
    (a) Distribution of $T_Q$ and $T_I$. The blue dashed line is the 
    fitted linear function of $T_Q$ and $T_I$. 
    Red dashed line is the cut limit for inner layer events; 
    (b) Histogram of event linearity index $L$, and the Gaussian fit of 
    linear (blue line) and nonlinear (red line) events; 
    (c) $T_Q$ Histogram for nonlinear events selected by 
    $L$ cut in (b). 
    The black dashed lines in (b) and (c) are the cut limit
    for inner layer events.
    }
\end{figure*}

The $T_Q$ and $T_I$ distribution of SSEs demonstrates two types of events: 
events gathered in a rodlike region in Fig.\ref{fig:linearity}(a) 
are referred to as linear events, and other events gathered in a cluster are 
referred to as nonlinear events. As shown in Fig.\ref{fig:linearity}, 
the charge drift time ($T_Q$) and a linearity index ($L$) are used to discriminate 
the linear and nonlinear events. The linearity index is defined as:

\begin{equation}
    L=T_I-\left(k\times T_Q+b\right),
    \label{eq:1}
\end{equation}

\noindent where fit parameters $k$ and $b$ are calculated via fitting 
$T_Q$ and $T_I$ of typical linear events with the function ($T_I=k\times T_Q+b$).
First, initial values of fit parameters ($k_0$ and $b_0$) are calculated by 
fitting events with $T_Q$ and $T_I$ below 500 ns. 
Then events with linearity $L=T_I-(k_0\times T_Q+b_0)$ in 
[-50, 50] ns are fitted to give the final value of $k$ and $b$. 
As shown in Fig.\ref{fig:linearity}(b), 
the distribution of linearity index $L$ is fitted with two Gaussian functions 
corresponding to linear and nonlinear events, respectively. 
The cut limit is set to ($\mu_{L,linear}-3\sigma_{L,linear}$), 
where $\mu_{L,linear}$ and $\sigma_{L,linear}$ are the mean and standard deviation 
of $L$ distribution for linear events. The distribution of $T_Q$ for nonlinear events selected 
by linearity index $L$ is fitted with a Gaussian function, and the cut limit is set to 
($\mu_{T,linear}-3\sigma_{T,linear}$), where $\mu_{T,linear}$ and $\sigma_{T,linear}$ 
are the mean and standard deviation of $T_Q$ distribution for nonlinear events as shown in Fig.\ref{fig:linearity}(c). 
The red dashed line in Fig.\ref{fig:linearity}(a) shows the discrimination limit set by 
the linearity index $L$ and the charge drift time $T_Q$.

\begin{figure*}[htbp]
    \centering
    \includegraphics
    [width=1.0\hsize]
    {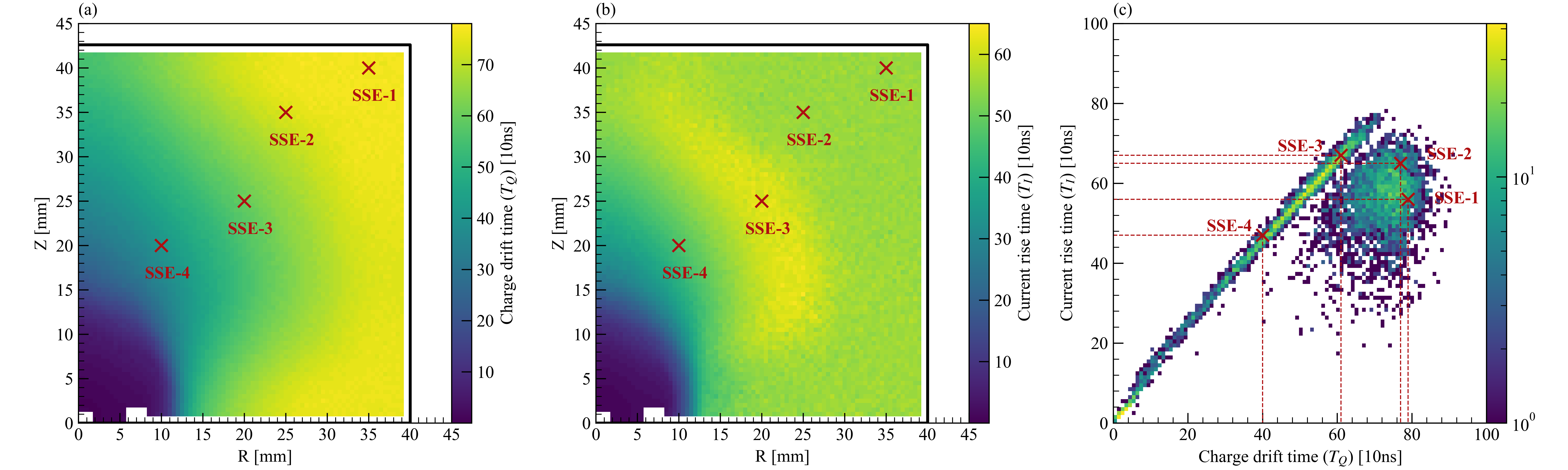}
    \caption{\label{fig:pss} 
    Pulse shape simulation for SSEs in different positions of the detector. 
    (a) Charge drift time ($T_Q$) for SSE as a function of the interaction position; 
    (b) Current rise time ($T_I$) for SSEs as a function of the interaction position;
    (c) Distribution of $T_Q$ and $T_I$ for pulses in (a) and (b), 
    those events are gathered in two clusters with a linear and nonlinear 
    relationship between $T_Q$ and $T_I$. 
    Red crosses mark the positions of four selected SSEs.
    }
\end{figure*}

\section{Detector segmentation model}\label{sec4}

\subsection{Demonstration of spatial distribution of 
linear and nonlinear events via pulse shape simulation}\label{sec4.1}

We perform a pulse shape simulation (PSS) for the HPGe 
detector to demonstrate the spatial distribution of the 
linear and nonlinear events. The electric field 
and weight potential field in the detector are calculated using the 
$mjd\_fieldgen$ package \cite{bib:19},
assuming a linear impurity profile in the Z-direction with an impurity density 
of $3.7\times 10^9$ $\rm{cm^3}$ and $8.0\times 10^9$ $\rm{cm^3}$ at the top 
and bottom surface of the crystal. SSEs with 1 MeV energy deposition 
are placed at different positions in the crystal. 
The corresponding charge pulses are calculated via the 
SAGE-PSS package \cite{bib:20}
and added with electric noise extracted from measured pulses.

Fig.\ref{fig:pss} demonstrates the $T_Q$ and $T_I$ as a function of 
the interaction position. As shown in Fig.\ref{fig:pss}(a) and (b), 
SSEs close to the $p$+ contact have shorter $T_Q$ and $T_I$. 
With the distance to contact increasing, the $T_Q$ and $T_I$ of induced 
pulses increase simultaneously, for instance, the SSE-3 and SSE-4. 
These events are typical linear events in Fig.\ref{fig:pss}(c). 
However, when SSEs near the top and side surfaces of the detector,  
their $T_Q$ and $T_I$ are not sensitive to their positions. 
Those SSEs, such as SSE-1 and SSE-2 are typical nonlinear events.
It can be explained by the Schockley-Ramo theory \cite{bib:25}:
when SSEs deposit energy near the outer surface of the detector,
the induced charge and current pulses will not exceed the 0.2\% 
of their maximum amplitude as charge carriers drift in the weak electric and 
weight potential field area near the surface. 
Thereby, the $T_Q$ and $T_I$ of those SSEs are not sensitive to the 
energy deposition position.

\subsection{Parameterized segmentation model}\label{sec4.2}

According to the pulse shape simulation, the linearity between $T_{Q}$ and $T_{I}$
of the SSE can be use to infer its hit position. 
We segment the detector into two layers referring to the positions of linear and nonlinear SSEs.
The boundary between the two layers is related to the electric and weight potential field of the detector.
And due to the lack of precise knowledge of the impurity profile within the Ge crystal, 
we can't rely on the PSS to calculate the shape of the two layers but take it as a reference.
Therefore, we take an empirical approach to build a segmentation model with 14 parameters
to described the boundary.

\begin{figure}[!htb]
    \centering
    \includegraphics
    [width=1.0\hsize]
    {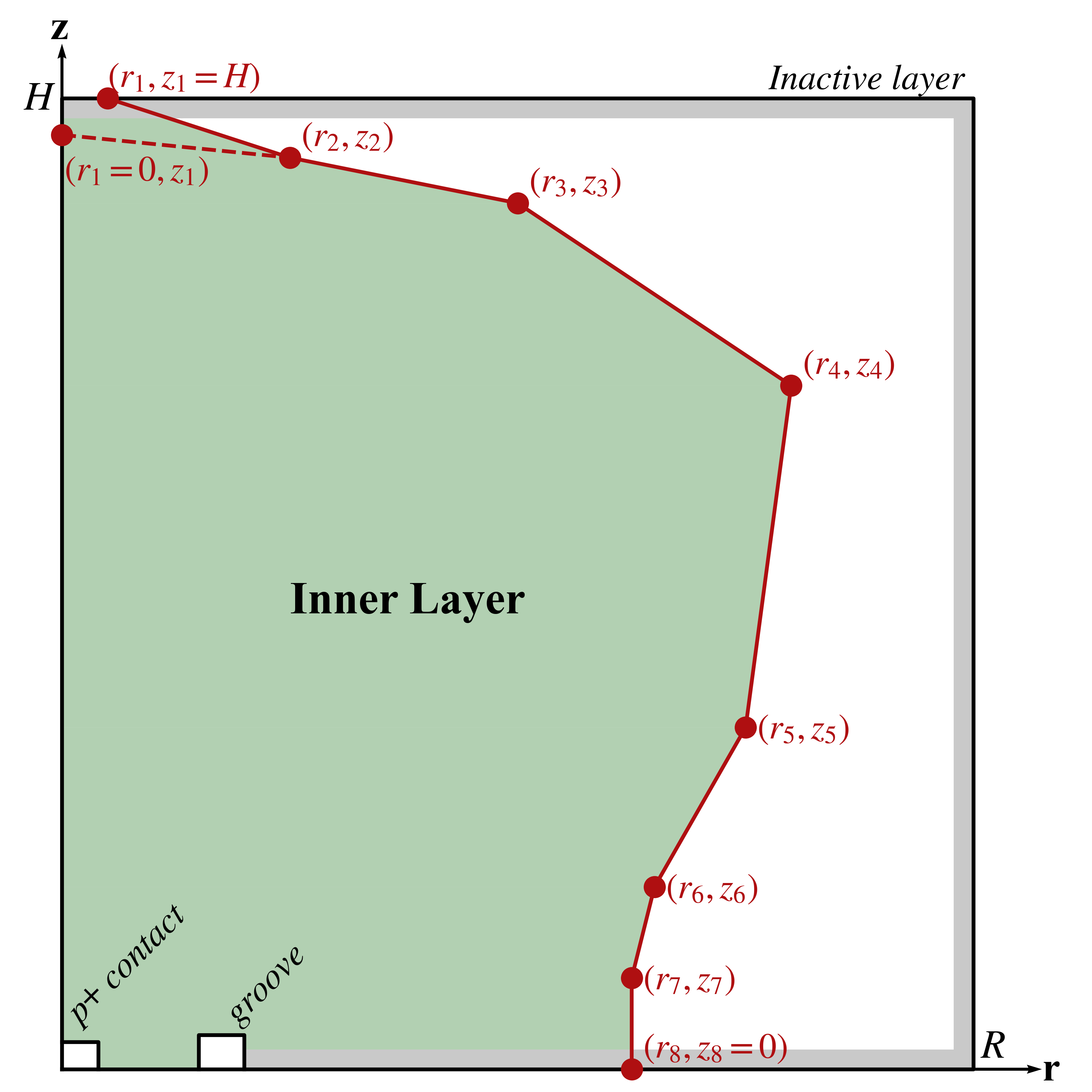}% Here is how to import EPS art
    \caption{\label{fig:model} 
    Parameterized segmentation model of the detector, 
    where $H$ and $R$ are the height and radius of the crystal. 
    The top spatial point $(r_1, z_1)$ could be on the top surface 
    $(z_1=H)$ or on the central axis $(r_1=0)$ of the crystal. 
    The green shadow region is the inner layer in the segmentation 
    model, and the gray shadow is the inactive layer in the $n$+ surface.}
\end{figure}

As shown in Fig.\ref{fig:model},
the boundary of the inner layer is the linear connection
of 8 spatial points.
It is worth noting that the number of spatial points in the model is arbitrary,
and it will be demonstrated later that the 8 points model is sufficient for this study.
Table.\ref{tab:model} lists the bound for each model parameter. 
As the model only requires the two layers to be continuous,
the first spatial opint $(r_{1},z_{1})$ could be on the top surface
or the central axis.
To determine the value of each model parameter,
we design and conduct a Th-228 scanning experiment.

\begin{table}
    \caption{Bounds for segmentation model parameters, 
    $R$ and $H$ are the radius and height of the Ge crystal.}\label{tab:model}%
    \renewcommand\arraystretch{1.5}
    \begin{tabular}{ccc}
    \toprule
    \multirow{2}{*}{Parameter}&\multirow{2}{*}{Parameter bound}\\
    &&\\ \hline
    \multirow{2}{*}{$(r_1,z_1)$}&$r_1=0$, $0<z_1<H$\\
    &or $z_1=H$, $0<r_1<R$\\
    $(r_2,z_2)$ & $r_1\leq r_2$, $z_2\leq z_1$\\
    $(r_3,z_3)$ & $r_2\leq r_3$, $z_3\leq z_2$\\
    $(r_4,z_4)$ & $r_3\leq r_4\leq R$, $z_4\leq z_3$\\
    $(r_5,z_5)$ & $r_5\leq R$, $z_5\leq z_4$\\
    $(r_6,z_6)$ & $r_6\leq r_5$, $z_6\leq z_5$\\
    $(r_7,z_7)$ & $r_7\leq r_6$, $z_7\leq z_6$\\
    $(r_8,z_8)$ & $0\leq r_8\leq r_7$, $z_8=0$\\
    \botrule
    \end{tabular}
\end{table}

\begin{figure}[!htb]
    \centering
    \includegraphics
    [width=1.0\hsize]
    {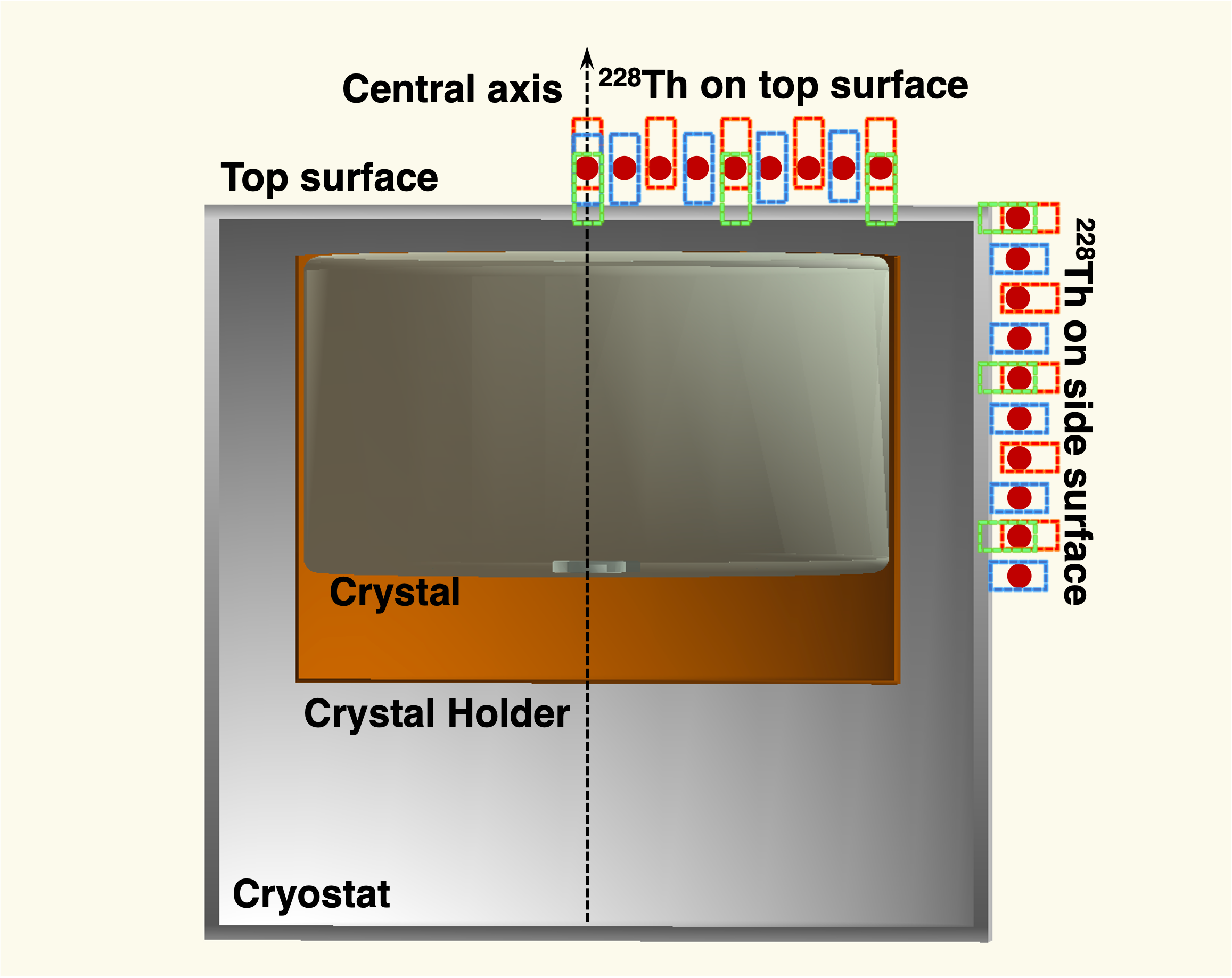}
    \caption{\label{fig:expTh228} 
    Schematic of Th-228 source positions in 
    calibration experiments. The red points 
    indicate the position of the Th-228 source. 
    The red, blue, and green dashed boxes mark the selected measurements for
    sub-datasets in the uncertainty assessment.
    The Th-228 source is mounted on a source holder. 
    The carbon fiber vacuum cryostat and the copper crystal holder are also shown.
    }
\end{figure}

\section{Optimization of segmentation model parameters}\label{sec5}

\subsection{Th-228 source scanning experiment}\label{sec5.1}

A Th-228 source is used to perform a scan of the detector top and side surfaces 
at 19 different positions as shown in Fig.\ref{fig:expTh228}.
A background measurement is also conducted for the detector.

\begin{figure}[!htb]
    \centering
    \includegraphics
    [width=1\hsize]
    {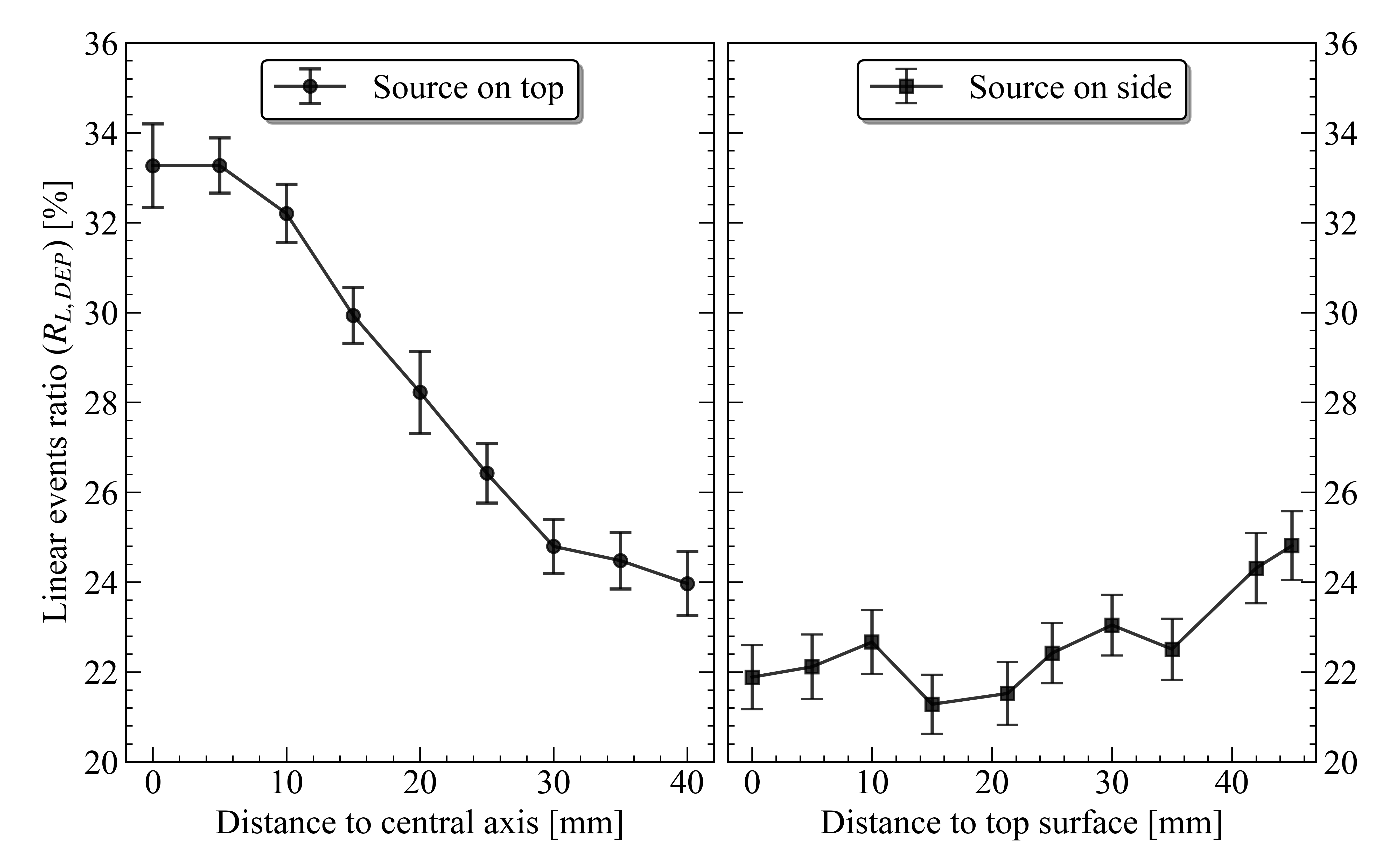}
    \caption{\label{fig:drt1} 
    Ratio of the linear event in selected DEP 
    events as a function of Th-228 source positions. 
    Error bars indicate the 1$\sigma$ uncertainty.}
\end{figure}

Events in the DEP region (1592.5$\pm$5 keV)
are selected as SSE candidates. After removing MSEs by 
the A/E cut, the linear events in the remaining SSEs are 
selected using the method in Sec \ref{sec3.3}. 
The ratio of linear events from the Th-228 source 
($R_{L,DEP}$) is then calculated by:

\begin{equation}
    R_{L,DEP}=\frac{N_{L,S}-N_{L,B}\cdot t_S/t_B}
    {N_{T,S}-N_{T,B}\cdot t_S/t_B},
    \label{eq:2}
\end{equation}

\noindent where $N_{T,S}$ and $N_{T,B}$ are total numbers of selected 
single-site DEP events in Th-228 and background measurements, 
respectively. $N_{L,S}$ and $N_{L,B}$ are numbers of selected 
linear events. $t_S$ and $t_B$ are the live time of source 
and background measurements. The uncertainty of $R_{L,DEP}$ 
is calculated by propagating the Poisson uncertainties of 
event counts in Th-228 and background measurement 
through Eq.(\ref{eq:2}). 
Fig.\ref{fig:drt1} shows the linear event ratio of SSEs in the DEP 
region as a function of Th-228 source positions. 
The $R_{L,DEP}$ decreased from 33.3\% to 24.0\% as the source 
moved from the top center to the edge of the detector. 
About 2.9\% changes in $R_{L,DEP}$  is observed when moving the source 
along the detector side surface.

\subsection{Spatial distribution of DEP events}\label{sec5.2}

\begin{figure}[!htb]
    \centering
    \includegraphics
    [width=1.0\hsize]
    {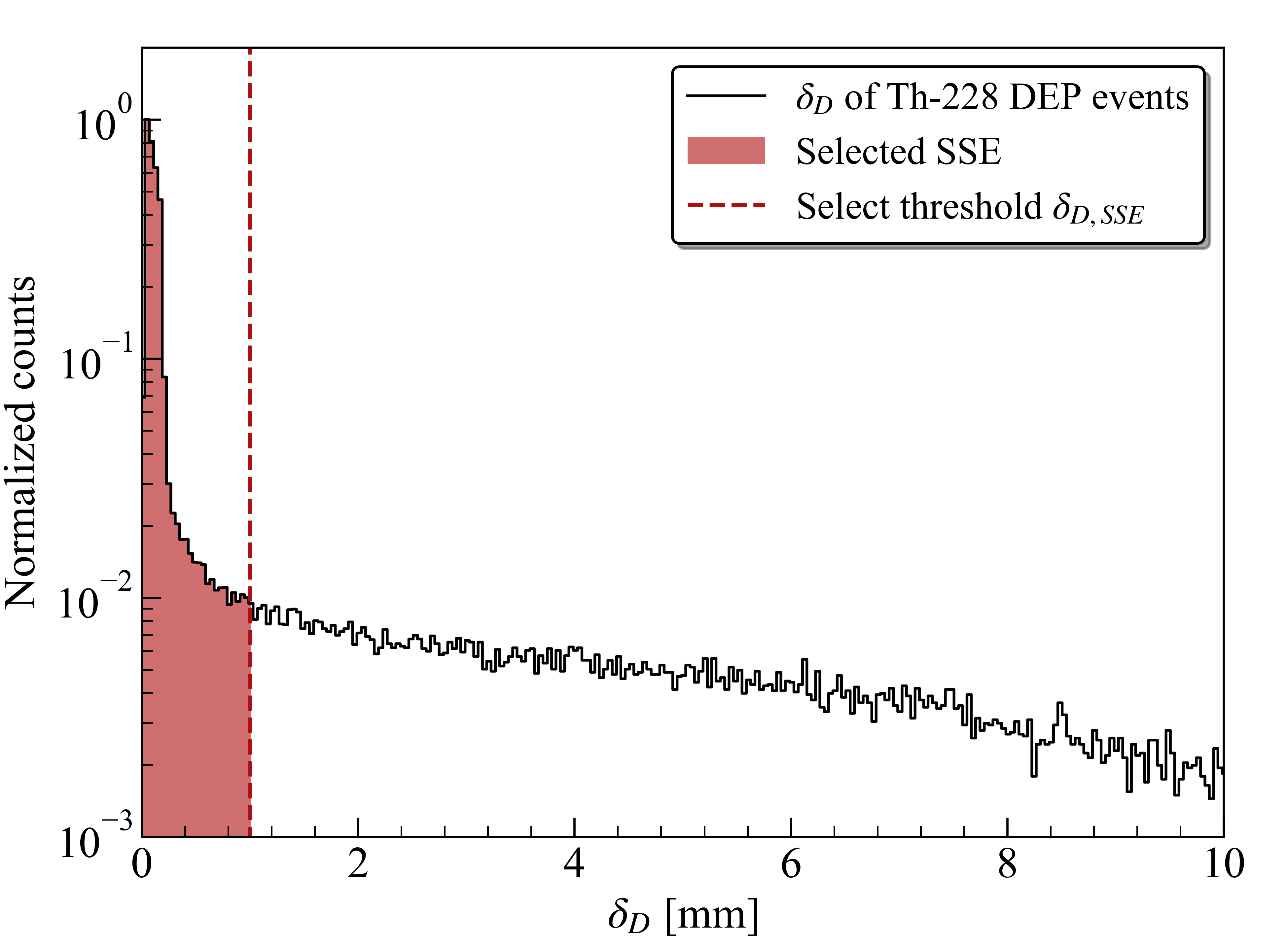}
    \caption{\label{fig:dD} 
    $\delta_D$ histogram for simulated DEP events with the 
    Th-228 source is placed at the center of the top detector surface. 
    }
\end{figure}

\begin{figure*}[hbtp]
    \centering
    \includegraphics
    [width=1.0\hsize]
    {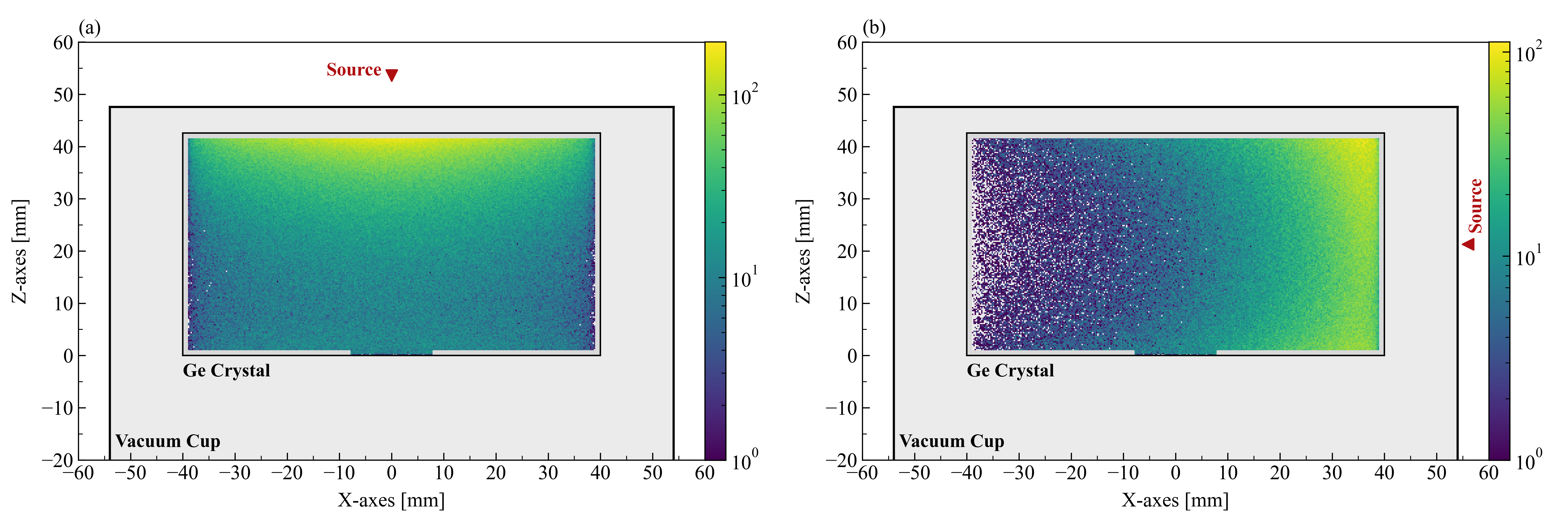}
    \caption{\label{fig:fDEP} 
    Spatial distribution of simulated SSEs in DEP region. 
    (a) Th-228 source in the center of the top surface; 
    (b) Th-228 source on the side of the detector. 
    The labels of the color bar represent the 
    distribution density (arbitrary unit).
    }
\end{figure*}

As the linear events are located in the inner layer of the segmentation 
model, the linear event ratio $R_{L,DEP}$ can be modeled by:

\begin{equation}
    R_{L,DEP}=\iint_{}^{}M(r,z\mid \theta)F_{DEP}(r,z)
    \cdot \rm{d}\it{r} \rm{d}\it{z},
    \label{eq:3}
\end{equation}

\begin{equation}
    M(r,z\mid \theta)=\left\{\begin{matrix}
       1 &(r,z)\in \rm inner\, layer\it \\ 
       0 & (r,z)\in \rm outer\, layer\it
       \end{matrix}\right.,
    \label{eq:4}
\end{equation}

\noindent where $M(r,z\!\mid\!\theta)$ is the select function for the 
inner event using the segmentation model, $\theta$ represents 
the model parameters in Table.\ref{tab:model}, 
$F_{DEP}(r,z)$ is the spatial distribution of SSEs in the DEP region. 
The energy deposition of $\gamma$ emitted by the Th-228 
source is simulated by Geant4 \cite{bib:24}.
The energy depositions occured in the inactive layer of the detector are not 
recorded in the simulation. The single-site events are 
selected by the $\delta_D$ parameter. $\delta_D$ is the average 
distance between the energy deposition points to the 
charge center of the event:

\begin{equation}
    \delta_D=\frac{1}{n}\sum_{i=0}^{n}\sqrt
    {(x_i-\hat{x})^2+(y_i-\hat{y})^2+(z_i-\hat{z})^2},
    \label{eq:5}
\end{equation}
\begin{equation}
    \hat{x}=\sum_{i=0}^{n}x_i\frac{E_i}{E_{tot}},
    \space 
    \hat{y}=\sum_{i=0}^{n}y_i\frac{E_i}{E_{tot}},
    \space 
    \hat{z}=\sum_{i=0}^{n}z_i\frac{E_i}{E_{tot}},
    \label{eq:6}
\end{equation}

\noindent where $n$ is the number of steps in one event, 
$(x_i,y_i,z_i)$ and $E_i$ are the hit position and 
energy deposition of the i-th step. 
$(\hat{x},\hat{y},\hat{z})$ and $E_{tot}$ are the charge center and 
total energy deposition of the event. 
Events with $\delta_D<\delta_{D,SSE}$ are selected as SSEs, 
where $\delta_{D,SSE}$ is determined by matching the 
survival fraction of DEP events in simulation with 
that of the A/E cut in the experiment. 
Fig.\ref{fig:dD} demonstrates a typical
$\delta_D$ distribution of simulated DEP 
events when the Th-228 source is at the top center of 
the detector. The charge center of the selected SSE 
is then used to simulate the spatial distribution $F_{DEP}(r,z)$. 
Fig.\ref{fig:fDEP}
shows the simulated $F_{DEP}(r,z)$ for the 
Th-228 source at two different positions.

\subsection{Optimization of model parameters}\label{sec5.3}

\begin{figure}[!htb]
    \centering
    \includegraphics
    [width=1.0\hsize]
    {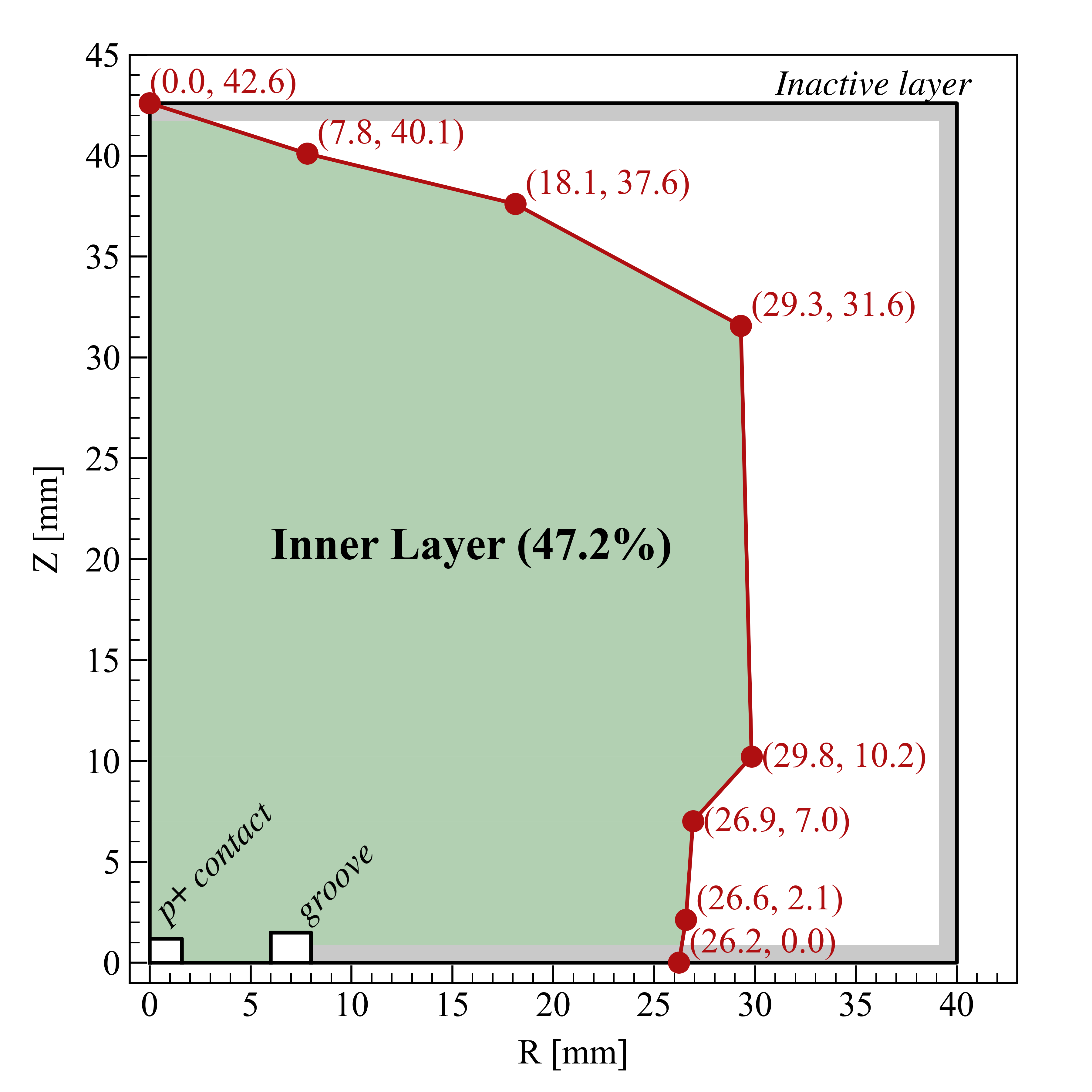}
    \caption{\label{fig:optM} 
    Optimized result of the segmentation model, 
    the volume of the inner layer is 47.2\% of 
    the total sensitive volume.
    }
\end{figure}

\begin{figure}[!htb]
    \centering
    \includegraphics
    [width=1.0\hsize]
    {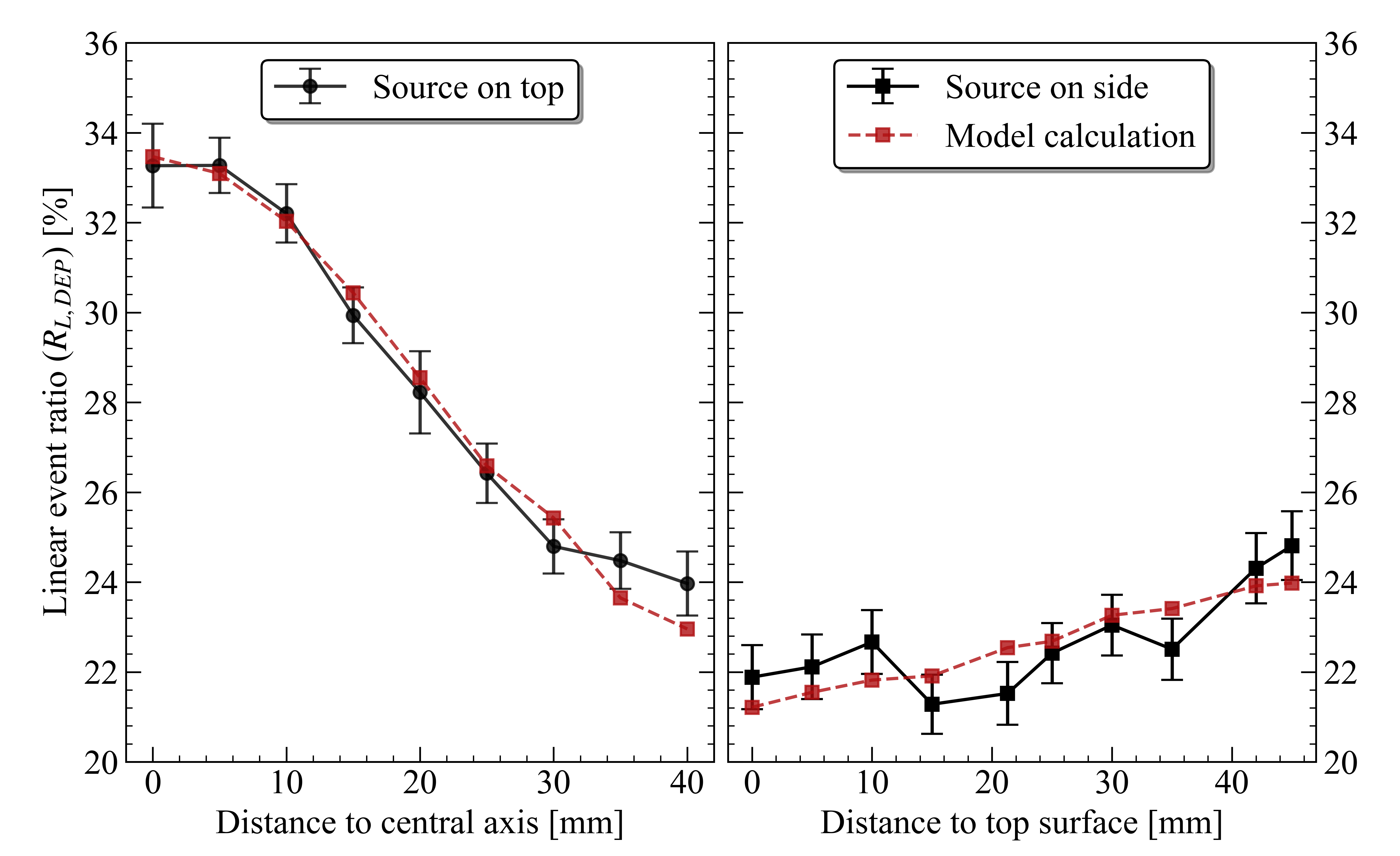}
    \caption{\label{fig:drt2} 
    Linear event ratio ($R_{L,DEP}$) calculated by 
    Eq.(\ref{eq:3}) using the optimized model 
    parameters (red squares). 
    The black dots and squares are the measured results from
    Fig.\ref{fig:drt1}.
    }
\end{figure}

As shown in Fig.\ref{fig:fDEP}, 
the position of the Th-228 source affects the spatial 
distribution of DEP events and therefore leads to  
different observed linear event ratios in Fig.\ref{fig:drt1}. 
Thus, we use a minimum-$\chi^2$ method to calculate the 
model parameters ($\theta$), in which $\chi^2$ is defined as:

\begin{equation}
    \chi^2=\sum_{k=1}^{19}
    \frac{\left (R_{k,exp}-\iint_{}^{}M(r,z\mid \theta)
    F_{DEP}(r,z)\rm d \it r \rm d \it z  \right )^2}{\sigma_k^2},
    \label{eq:7}
\end{equation}

\noindent where $R_{k,exp}$ is the measured linear event ratio for 
Th-228 source at position $k$ ($k$=1,2,$\dots$19), 
$\sigma_k$ is the corresponding uncertainty 
of $R_{k,exp}$. $F_{DEP,k}(r,z)$ is the simulated spatial 
distribution of single-site DEP events
for the Th-228 source at position $k$. 
The minimalization of $\chi^2$ is implemented by the 
genetic algorithm using a python-based calculation package 
Geatpy \cite{bib:22}. 
Fig.\ref{fig:optM} shows the optimized results. 
The volume of the inner layer is 47.2\% of the total 
sensitive volume of the detector. 
The linear event ratios calculated by Eq.\ref{eq:3} 
using the optimized model parameters
are shown in Fig.\ref{fig:drt2}.
The fit result agrees well with the measurements,
the $p-value$ of the $\chi^2$ fit is 0.701.

\section{Uncertainty assessment and model validation}\label{sec6}

\begin{figure*}[htbp]
    \centering
    \includegraphics
    [width=1.0\hsize]
    {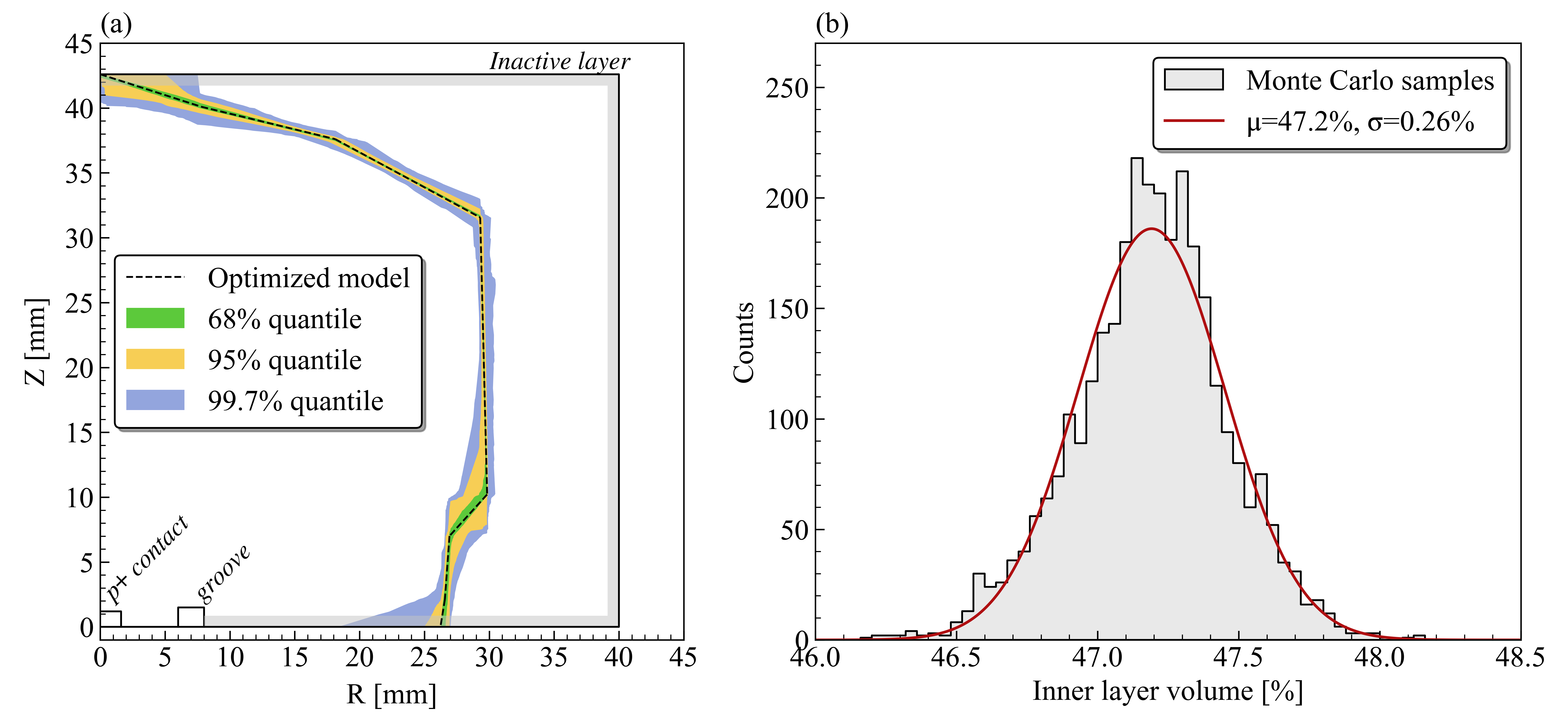}
    \caption{\label{fig:uncstat} 
    (a) Inner layer shapes of the 3000 Monte Carlo datasets. 
    The green, yellow, and blue shadow bands are corresponding to
    68\%, 95\%, and 99.7\% quantiles, respectively. 
    The gray shadow is the inactive layer on the $n$+ surface. 
    (b) Distribution of inner layer volumes. 
    The red line is the fit of inner layer volumes 
    using a Gaussian function, $\mu$ and $\sigma$ are 
    the mean and standard deviation, respectively.
    }
\end{figure*}

\begin{figure*}[htbp]
    \centering
    \includegraphics
    [width=1.0\hsize]
    {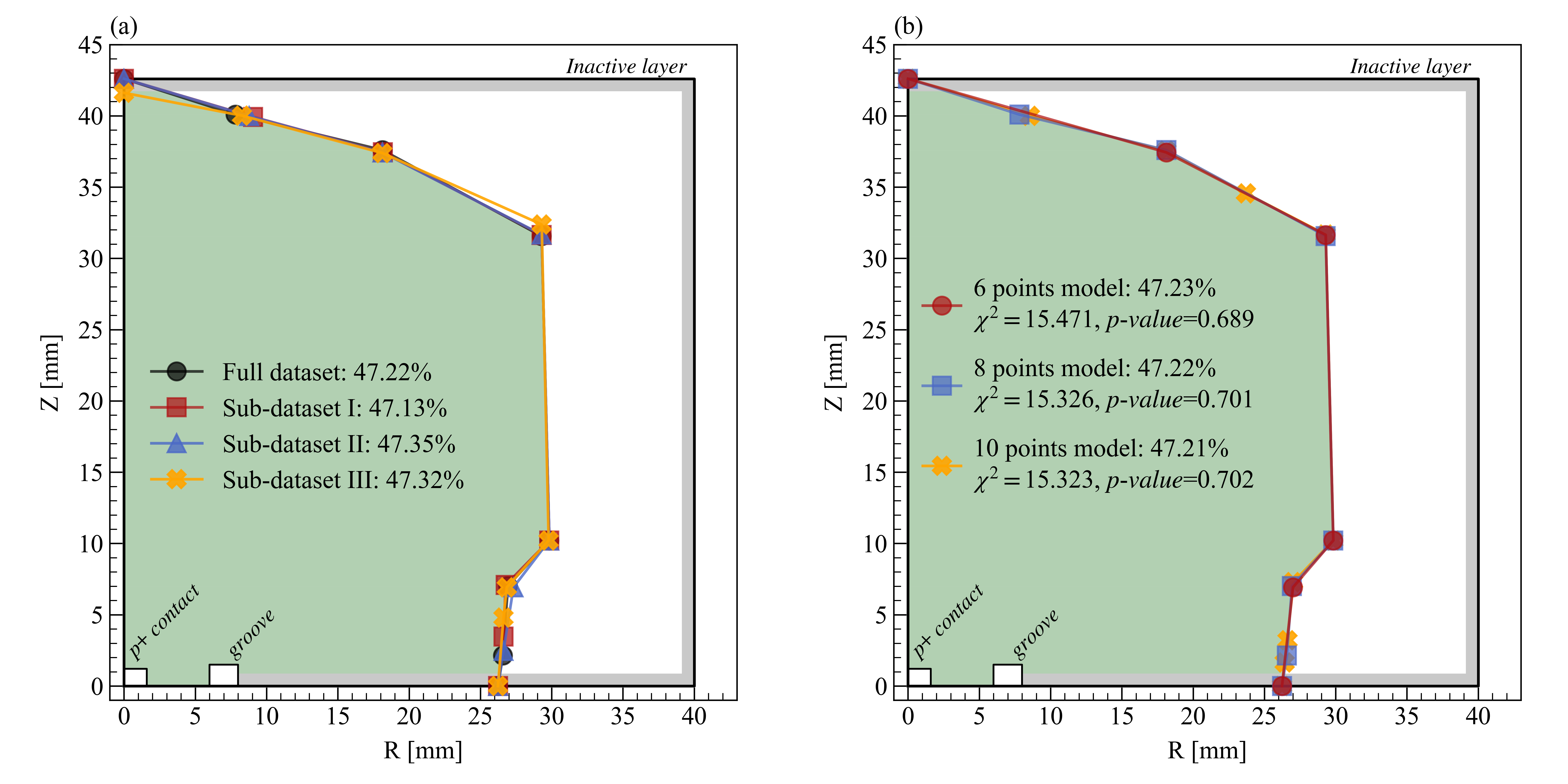}
    \caption{\label{fig:uncsys} 
    (a) Optimized results using different datasets, 
    full dataset (black line) consists of all measured data, 
    sub-dataset I, II, III are selected from the full 
    dataset. 
    (b) Optimized results for three different models, 
    the chi-square ($\chi^2$) and $p$-$value$ 
    are given to demonstrate the fit goodness of each model. 
    The gray shadow regions in both figures are the inactive layer 
    on the detector $n$+ surface.
    }
\end{figure*}

Uncertainties of the shape and volume 
of the inner layer in the optimized model 
mainly consist of three parts:
\begin{itemize}
    \item[(1)]
    Uncertainty of the linear event ratio 
    ($R_{L,DEP}$) propagated by the $\chi^2$-method 
    is evaluated using a toy Monte Carlo method.
    3000 Monte Carlo datasets are 
    generated assuming a Gaussian distribution for 
    the $R_{L,DEP}$ with the mean and standard deviation
    equal to the measured value and uncertainty, respectively.
    Model parameters are recalculated for each dataset 
    following the same analysis in Sec \ref{sec5.3}.
    The distribution of inner layer shapes and volumes 
    for the 3000 samples are illustrated in Fig.\ref{fig:uncstat}. 
    The distribution of inner layer volume is fitted with 
    a Gaussian function, and the standard deviation, $\pm$0.26\%,
    is adopted as the statistical uncertainty.
    \item[(2)]
    Systematic uncertainty due to the choice of dataset: 
    we divide the measured data in Fig.\ref{fig:drt1} 
    into three sub-datasets. 
    Sub-dataset I and II each consists of ten measured data 
    (marked by red dashed boxes for sub-dataset I, 
    and blue dashed boxes for sub-dataset II in Fig.\ref{fig:expTh228}).
    sub-dataset III consists of six measured data 
    (green dashed boxes in Fig.\ref{fig:expTh228}).
    The fitting of model parameters
    are performed in each sub-dataset, and the
    largest difference in inner 
    layer volume between all sub-datasets and the full dataset
    (Fig.16 (a)) is $\pm$0.22\% as a systematic uncertainty.
    \item[(3)]
    Systematic uncertainty due to the construction 
    of the segmentation model: we reconstruct the segmentation 
    model using 6 spatial points (10 free parameters) 
    and 10 spatial points (18 free parameters) and 
    calculate the model parameters using the full dataset. 
    Fig.\ref{fig:uncsys}(b) shows the optimized results for the reconstructed models. 
    The overall shape and volume of the inner layer are similar in the three models,
    and the largest difference in inner layer volume is 0.02\%,
    which is about 10 times smaller than the other two uncertainties and thereby negligible.
    This indicates the 8-point segmentation model is sufficient in this study.
\end{itemize}

Including the statistical and systematic uncertainties discussed above, 
the volume of the inner layer is given as 47.2\%$\pm$0.26\%(stat.)$\pm$0.22\%(sys.)

\begin{figure*}[htbp]
    \centering
    \includegraphics
    [width=1.0\hsize]
    {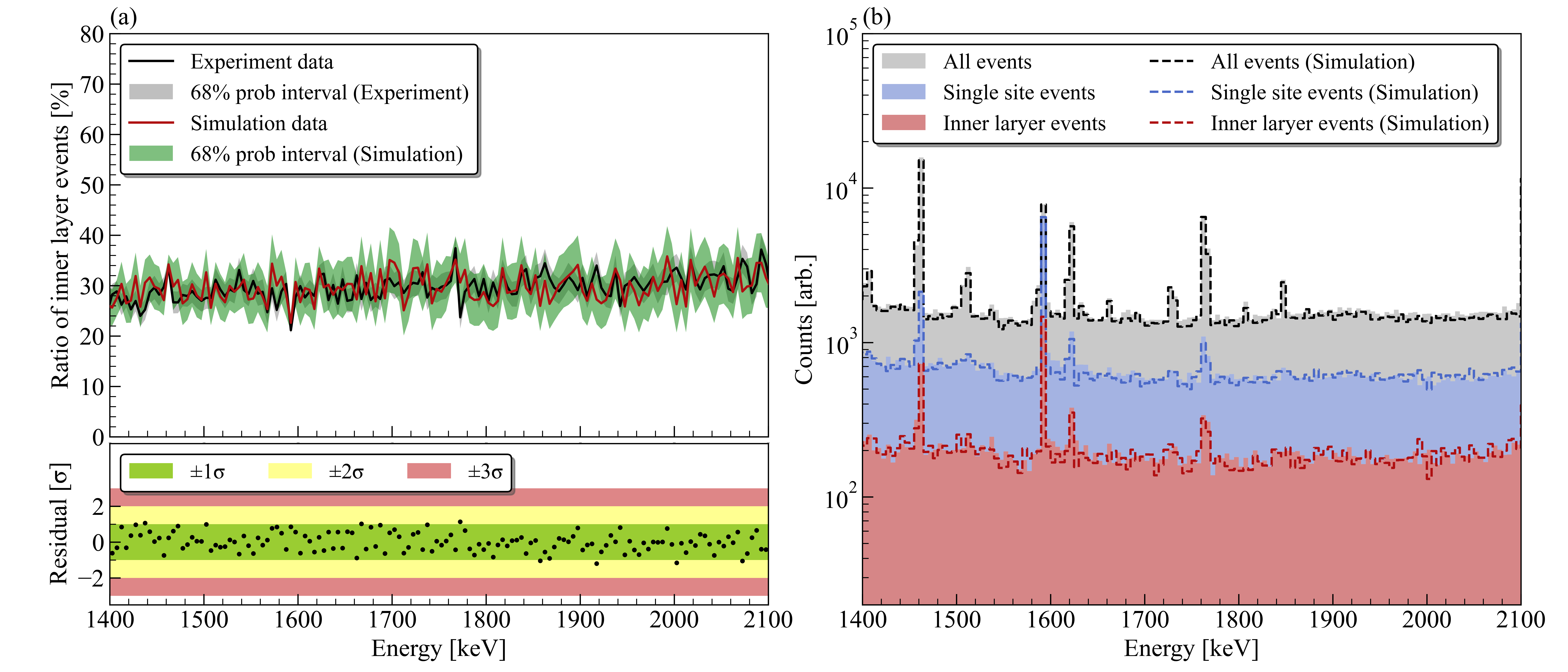}
    \caption{\label{fig:val} 
    Comparison of simulation and experiment for Th-228 
    source placed on the side of the detector. 
    (a) The linear event ratio as a function of energy, 
    The uncertainty band for simulation 
    (the green shadow) consists of uncertainty 
    from the inner layer shape 
    (68\% quantile region in Fig.\ref{fig:uncstat}(a)) 
    and statistical uncertainty in simulation.
    The normalized residuals are shown in the bottom figure, 
    (b) Measured and simulated spectra in 1400-2100 keV region. 
    }
\end{figure*}

The measured Th-228 spectra are compared to the simulated 
spectra to validate the segmentation model. 
The energy depositions of the $\gamma$-rays emitted 
from the Th-228 source are simulated via Geant4 and 
added with the energy resolution of the detector. 
The SSEs are selected using the $\delta_D$ parameter 
defined in Sec \ref{sec5.2}, and the inner layer events 
are selected by their charge center positions 
defined by Eq.(\ref{eq:6}). 
The measured background spectrum is scaled by the 
live time of measurement and added to the simulated spectra.

Fig.\ref{fig:val} compares the spectra and ratio of inner layer
SSE events between simulation 
and experimental results for one of Th-228 source measurements. 
The gray band in Fig.\ref{fig:val}(a) is
the statistic uncertainty of experiment data,
the green band is the combination of the statistic and 
systematic uncertainties in the simulation.
In this case, the systematic uncertainty is taken as the discrepancy between
linear event ratios corresponding to the innermost 
and outmost shape of the 68\% quantile of the inner layer 
(the green region in fig.\ref{fig:uncstat}(a)).
Fig.\ref{fig:val}(b) is the comparison of measured and simulated spectra,
it demonstrates that the $\delta_D$ cut in the simulation is a good approximation for the A/E cut, 
and the spectra of inner layer events also show a good agreement between 
the simulation and measurement in the 1400-2100 keV energy region.

\section{Background suppression performance of virtual segmentation}\label{sec7}

In the search for Ge-76 $0\nu\beta\beta$ decay using HPGe detectors,
backgrounds, mostly $\gamma$-rays and electrons from outside the detector,
have to penetrate the outer layer of the detector to deposit their energy in the inner layer.
Thus, the outer layer in the virtual segmentation could act as a shielding for the inner layer,
and a lower background level of the inner layer may improve the detection sensitivity.

\begin{figure}[!htb]
    \centering
    \includegraphics
    [width=1.0\hsize]
    {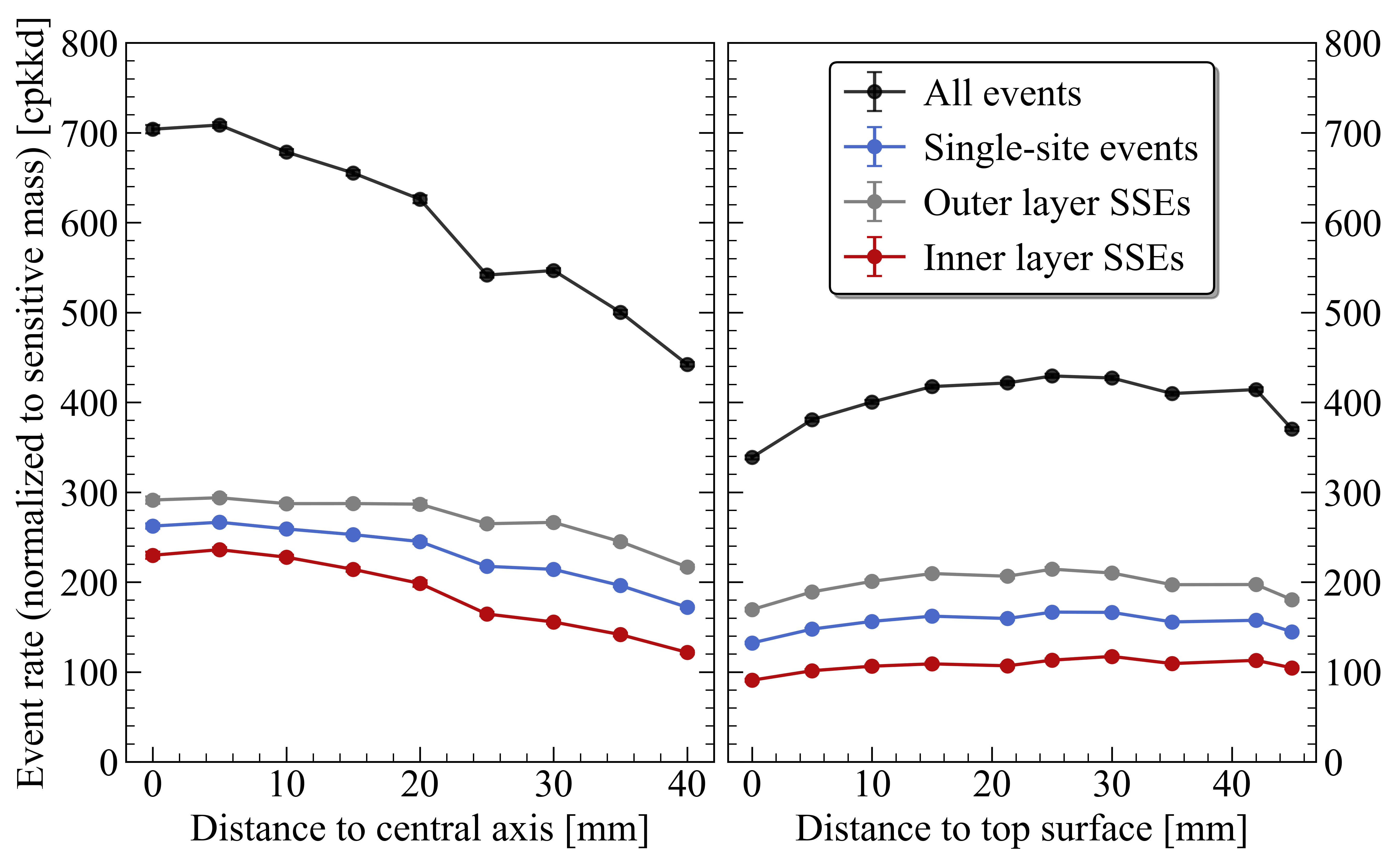}
    \caption{\label{fig:bkg1} 
    Measured spectra for the Th-228 source on the side 
    surface of the detector. 
    cpkkd represents counts per kg per keV per day,
    $Q_{\beta\beta}$ is the energy of Ge-76 $0\nu\beta\beta$ signal.
    }
\end{figure}

We use the Th-228 scanning data to evaluate the background suppression power
of the virtual segmentation.
The count rates in spectra are normalized to unit sensitive mass to include
the mass loss due to the analysis volume selection. 
The masses of the detector are 1.052 kg and 0.496 kg for the total sensitive volume and 
the inner layer, respectively.
Fig.\ref{fig:bkg1} demonstrates 
spectra before and after A/E cut and inner layer event 
selection when the Th-228 source is placed 
on the side of the detector. 
First the whole detector is selected as the analysis volume and
the A/E cut is applied to removes multi-site events (gray and blue regions in Fig.\ref{fig:bkg1}).
Then the inner layer of the virtual segmentations is selected as the analysis volume,
a further reduction on the event rate is shown in Fig.\ref{fig:bkg1} (red region).
It is expected that the SSEs mostly come from the
single Compton scattering of high energy $\gamma$-rays emitted from the source
and are clustered near the surface of the detector.
Thereby the inner layer has a lower background level in the detector.

\begin{figure}[!htb]
    \centering
    \includegraphics
    [width=1\hsize]
    {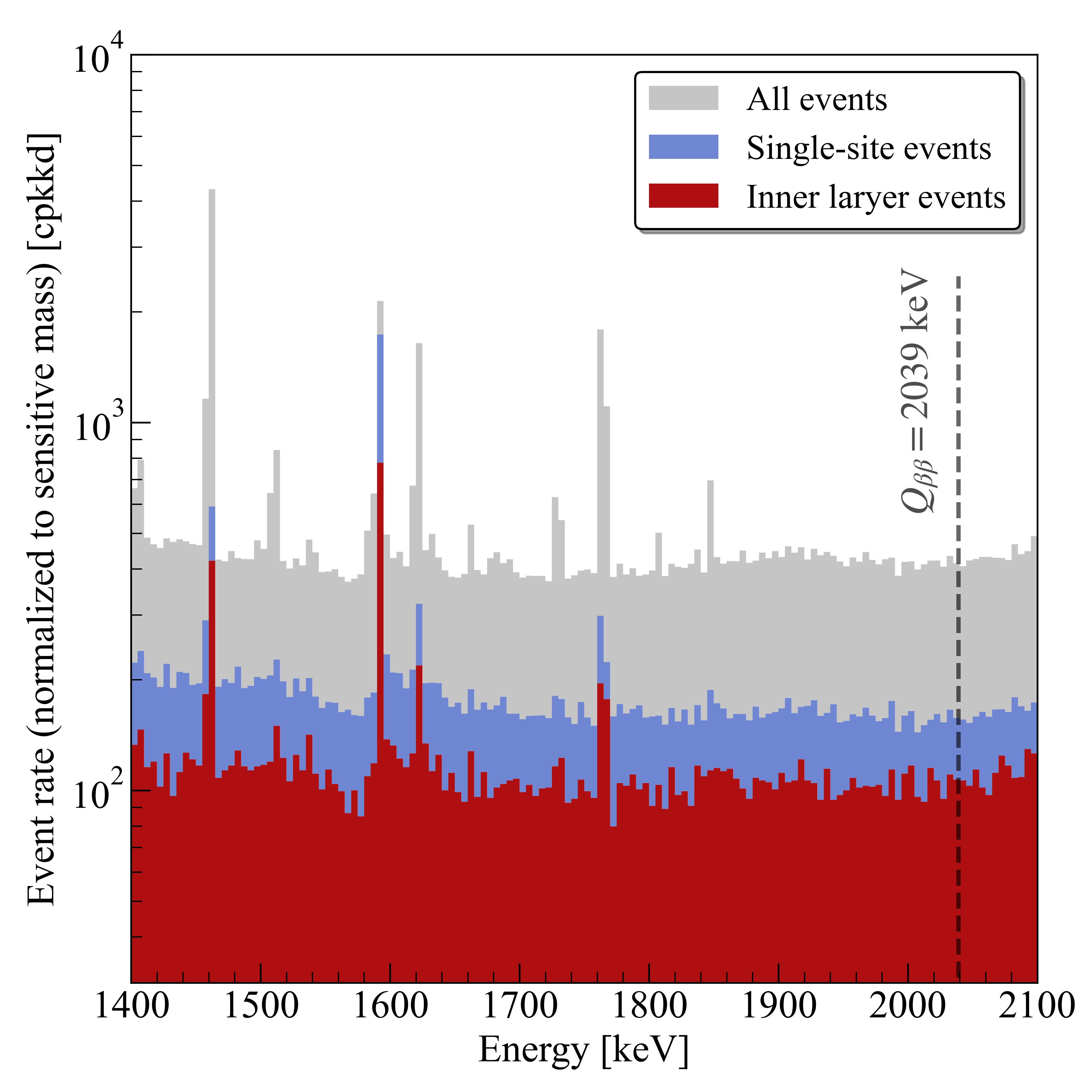}
    \caption{\label{fig:bkg2} 
    Event rate in $0\nu\beta\beta$ signal region (1900-2100 keV)
    as a function of Th-228 source position. 
    The left and right figures show the event rate for the Th-228 source placed
    on the top and side surface of the detector, respectively.
    }
\end{figure}

Fig.\ref{fig:bkg2} shows the event rate in the
$0\nu\beta\beta$ signal region (1900-2100 keV)
as a function of the Th-228 source positions. 
The highest background suppression power 
is achieved when the Th-228 source is at the side of the detector.
In this case, the A/E cut reduces the event rate by 62\%, 
and the virtual segmentation yeilds a further reduction of 12\% on the basis of the A/E cut.

In future $0\nu\beta\beta$ experiments using 
small contact HPGe detectors, this method might be used to
further suppress background in the signal region. 
Especially for experiments using a liquid argon (LAr) veto system 
where the HPGe detector is directly immersed in LAr, 
such as GERDA \cite{bib:1},
the LEGEND \cite{bib:23}, 
and CDEX-300$\nu$ experiments \cite{bib:3}.
The background from K-42 (daughter of cosmogenic Ar-42 in LAr) 
beta-decay is mainly located in the surface of the detector, 
therefore might be suppressed if the inner layer is selected as the analysis volume.
It should be noted that the balance between a lower background and the loss in detector sensitive mass 
should be considered in the the searching for the $0\nu\beta\beta$ signal.

Furthermore, the discrepancy between the inner and outer layer SSE spectrum 
could be used to infer the location of the background source.
A more precise background model could be built by 
fitting the spectra of events in the inner and the outer layer simultaneously.

\section{Summary}\label{sec8}
In this study, 
we develop a virtual segmentation model for a small contact HPGe detector
and demonstrate its background suppression capability in the Ge-76 $0\nu\beta\beta$ signal region.
The HPGe detector is virtually segmented into two layers, and
a selection algorithm based on charge pulse drift time ($T_{Q}$) and current rise time ($T_{I}$)
is established to identify the position of the single-site event. 
The shape and volume of the inner layer in the segmentation model are determined
using the DEP events in a series of Th-228 source calibration experiments. 
The volume of the inner layer is evaluated to be 
47.2\%$\pm$0.26(stat.)\%$\pm$0.22(sys.)\% 
of the total sensitive volume of the detector.

The background suppression power of the virtual segmentation
in Ge-76 $0\nu\beta\beta$ signal region
is evaluated by the Th-228 scanning data.
Choosing the inner layer as the analysis volume, a further 12\% reduction of 
background is achieved when the Th-228 source is on the side of the detector. 
Other backgrounds in the $0\nu\beta\beta$ signal region, 
especially those clustered on the surface of the detector, 
such as Ar-42 in future $0\nu\beta\beta$ experiments,
could also be reduced by the virtual segmentation.

The principle of the virtual segmentation can be extended to other 
small contact HPGe detectors, for instance, point-contact Ge (PCGe) and broad energy Ge (BEGe) detectors.

\section*{Acknowledgments}
This work was supported by
the National Key Research and Development Program of China 
(Grant No. 2022YFA1604701) and 
the National Natural Science Foundation of China 
(Grants No. 12175112).
We would like to thank CJPL and its staff for supporting this work. 
CJPL is jointly operated by Tsinghua University and Yalong River Hydropower Development Company.

\end{document}